\documentclass[10pt]{article}

\usepackage{graphicx}
\usepackage{dcolumn}
\usepackage{bm}

\begin{document}
\begin{center}
{{\LARGE \bf Relativistic hydrodynamics from Boltzmann equation with modified collision term}} \\
\bigskip
{\large \bf Amaresh Jaiswal\footnote{email: amaresh@tifr.res.in}, Rajeev S. Bhalerao and Subrata Pal } \\
{Tata Institute of Fundamental Research, Mumbai, India} 
\bigskip
\end{center}

\begin{abstract}
Generalizing the collision term in the relativistic Boltzmann 
equation to include nonlocal effects, and using Grad's 14-moment 
approximation for the single-particle distribution function, we 
derive evolution equations for the relativistic dissipative fluid 
dynamics and compare them with the corresponding equations obtained 
in the standard Israel-Stewart and related approaches. Significance 
of this generalization on hydrodynamic evolution is demonstrated in 
the framework of one-dimensional scaling expansion.
\end{abstract}


\section{Introduction}
Relativistic dissipative hydrodynamics has been quite successful in 
explaining the spectra and azimuthal anisotropy of particles 
produced in heavy-ion collisions at the RHIC and LHC \cite 
{Song:2010mg}. Apart from its applications, theoretical formulation 
of relativistic dissipative hydrodynamics is quite a challenging 
task by itself. The first-order dissipative fluid dynamics commonly 
known as relativistic Navier-Stokes (NS) theory involves parabolic 
differential equations and suffers from acausality and instability. 
The second-order or Israel-Stewart (IS) theory \cite {Israel:1979wp} 
with its hyperbolic equations restores causality but may not 
guarantee stability. The correct formulation of relativistic 
dissipative fluid dynamics is still unresolved and is currently 
under intense investigation \cite{Denicol:2010xn,Jaiswal:2012qm}. 

It is essential to note that all formulations of the second-order 
dissipative hydrodynamics employing the Boltzmann equation (BE) make 
a strict assumption of locality in the configuration space for the 
collision term \cite{Israel:1979wp}. This means, the collisions that 
increase or decrease the number of particles with a given momentum 
$p$, in an infinitesimal space-time volume element, are assumed to 
occur at the same point $x^\mu$. This makes the collision integral a 
purely local functional of the single-particle phase-space 
distribution function $f(x,p)$ independent of the derivatives 
$\partial^\mu f$. Although $f(x,p)$ may not vary significantly over 
the length scale of a single collision event, its variation over the 
length scales extending over many inter-particle spacings may not be 
negligible. Including the gradients of $f(x,p)$ in the collision 
term gives rise to different evolution equations for the dissipative 
quantities.

In this article, we provide a new formulation of the dissipative 
hydrodynamic equations within kinetic theory by using a nonlocal 
collision term in the Boltzmann equation: $p^\mu \partial_\mu f = 
C[f]$.

\section{Non-locality in collision term}
\begin{figure}
\begin{center}
\includegraphics[scale=0.18]{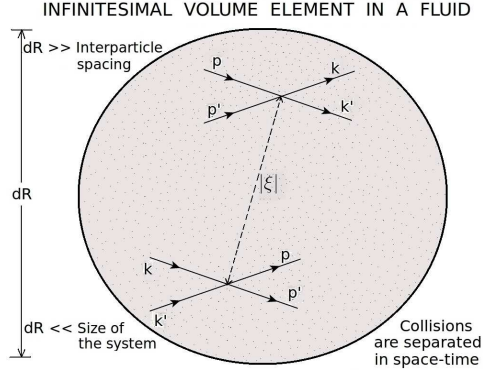}
\caption{Collisions $kk' \to pp'$ and $pp' \to kk'$ separated 
		 by a spacetime interval $\xi^\mu$ within an infinitesimal 
		 fluid element of size $dR$, containing a large number of 
		 particles represented by dots.}		 
\label{flelmnt}
\end{center}
\end{figure}

For two-body elastic collisions, the collision term is 
\begin{equation}\label{coll} 
C[f]= \frac{1}{2} \int dp' dk \ dk' \  W_{pp' \to kk'}
(f_k f_{k'} \tilde f_p \tilde f_{p'} - f_p f_{p'} \tilde f_k \tilde f_{k'}), 
\end{equation} 
where $ W_{pp' \to kk'}$ is the collisional transition rate, $f_p 
\equiv f(x,p)$, $\tilde f_p \equiv 1-r f(x,p)$ with $r = 1,-1,0$ for 
Fermi, Bose, and Boltzmann gas and $dp = d{\bf p}/[(2 \pi)^3E_{\bf 
p}]$.  The first and second terms in Eq. (\ref {coll}) refer to the 
processes $kk' \to pp'$ and $pp' \to kk'$, respectively. These 
processes are traditionally assumed to occur at the same space-time 
point $x^\mu$ with an underlying assumption that $f(x,p)$ is 
constant not only over a region characterizing a single collision, 
but also over an infinitesimal fluid element of size $dR$, large 
compared to the interparticle separation. It is important to note 
that the space-time points at which the above two processes occur 
may be separated by a small interval $\xi^\mu$ within $d^4R$ (see 
Fig. \ref{flelmnt}). With this realistic viewpoint, the second term 
in Eq. (\ref{coll}) involves $f(x-\xi,p)f(x-\xi,p')\tilde 
f(x-\xi,k)\tilde f(x-\xi,k')$, which on Taylor expansion up to 
second order in $\xi^\mu$, results in the modified BE 
\cite{Jaiswal:2012qm}
\begin{equation}\label{MBE}
p^\mu \partial_\mu f = C[f] + \partial_\mu(A^\mu f) + \partial_\mu\partial_\nu(B^{\mu\nu}f).
\end{equation}
The momentum dependence of coefficients,
\begin{eqnarray}\label{coeff1}
A^\mu &=& \frac{1}{2} \int dp' dk \ dk' \ \xi^\mu W_{pp' \to kk'}
f_{p'} \tilde f_k \tilde f_{k'}, \nonumber \\
B^{\mu\nu} &=& -\frac{1}{4} \int dp' dk \ dk' \ \xi^\mu \xi^\nu W_{pp' \to kk'}
f_{p'} \tilde f_k \tilde f_{k'},
\end{eqnarray}
can be made explicit by expressing them in terms of the available
tensors $p^\mu$ and the metric $g^{\mu\nu} \equiv {\rm
  diag}(1,-1,-1,-1)$ as $A^\mu = ap^\mu$ and $B^{\mu\nu}=
b_1g^{\mu\nu} + b_2 p^\mu p^\nu$. The scalar coefficients $a$, $b_1$
and $b_2$ are functions of $x^\mu$.

The conserved particle current, $N^\mu$, and the energy-momentum 
tensor, $T^{\mu\nu}$ have the standard form \cite{Israel:1979wp}:
\begin{eqnarray}\label{NTD}
N^\mu \!\!\!&=&\!\!\! \int dp \ p^\mu f = nu^\mu + n^\mu,  \nonumber\\
T^{\mu\nu} \!\!\!&=&\!\!\! \int dp \ p^\mu p^\nu f = \epsilon u^\mu u^\nu
-(P+\Pi)\Delta ^{\mu \nu} + \pi^{\mu\nu}.
\end{eqnarray}
Conservation of current, $\partial_\mu N^\mu=0$ and energy-momentum tensor, 
$\partial_\mu T^{\mu\nu} =0$, yield the fundamental evolution equations for 
$n$, $\epsilon$ and $u^\mu$
\begin{eqnarray}\label{evol}
Dn+n\partial_\mu u^\mu + \partial_\mu n^\mu \!\!&=&\!\! 0, \nonumber \\
D\epsilon + (\epsilon+P+\Pi)\partial_\mu u^\mu - \pi^{\mu\nu}\nabla_{(\mu} u_{\nu)} \!\!&=&\!\! 0,  \nonumber\\
(\epsilon+P+\Pi)D u^\alpha - \nabla^\alpha (P+\Pi) + \Delta^\alpha_\nu \partial_\mu \pi^{\mu\nu}  \!\!&=&\!\! 0.
\end{eqnarray}
Conservation of current and energy-momentum implies vanishing 
zeroth and first moments of the collision term $C_m[f]$ leading to 
three constraint equations for the coefficients ($a, b_1, b_2$) \cite
{Jaiswal:2012qm},
\begin{eqnarray}\label{param}
\partial_\mu a = 0, \quad\quad
\partial^2\left(b_1 a_{00} \right) 
+ \partial_\mu \partial_\nu\left( b_2 I^{\mu\nu}\right) \!\!&=&\!\! 0, \nonumber\\
u_\alpha \partial_\mu \partial_\nu \left( b_2 I^{\mu\nu\alpha} \right)  
+ u_\alpha \partial^2 \left(b_1 n u^\alpha \right) \!\!&=&\!\! 0.
\end{eqnarray}
In order to obtain the second-order evolution equations for dissipative 
quantities, we consider second moment of the modified BE, Eq. (\ref{MBE})
\begin{equation}\label{BE2}
\int\!\! dp \  p^\alpha p^\beta p^\gamma \partial_\gamma f 
= \!\!\int\!\! dp \ p^\alpha p^\beta \big[ C[f]
+ p^\gamma \partial_\gamma(af) + \partial^2(b_1f_0)  + (p \cdot \partial)^2 (b_2f_0) \big]. 
\end{equation}
Using Grad's 14-moment approximation for the single particle 
distribution function in orthogonal basis, 
\begin{eqnarray}\label{G14}
f = f_0 + f_0 \tilde f_0 \left( \lambda_\Pi \Pi + \lambda_n n_\alpha p^\alpha 
+ \lambda_\pi \pi_{\alpha\beta} p^\alpha p^\beta \right),
\end{eqnarray}
we finally obtain the following evolution equations for the dissipative
fluxes
\begin{eqnarray}
\Pi \!\!\!\!&=&\!\!\!\! \tilde a \Pi_{\rm NS} 
- \beta_{\dot \Pi} \tau_\Pi \dot \Pi
+ \tau_{\Pi n} n \cdot \dot u - l_{\Pi n} \partial \cdot n
- \delta_{\Pi\Pi} \Pi\theta
+ \lambda_{\Pi n} n \cdot \nabla \alpha  \nonumber\\
&&\!\!\!\!
+ \lambda_{\Pi\pi} \pi_{\mu\nu} \sigma^{\mu\nu}
+ \Lambda_{\Pi\dot u} \dot u \cdot \dot u
+ \Lambda_{\Pi\omega} \omega_{\mu\nu} \omega^{\nu\mu} + (8 \ {\rm terms}) , \label{bulk}\\
n^\mu \!\!\!\!&=&\!\!\!\! \tilde a n^\mu_{\rm NS}
- \beta_{\dot n} \tau_n \dot n^{\langle \mu \rangle}
+ \lambda_{nn} n_\nu \omega^{\nu\mu}
- \delta_{nn} n^\mu \theta
+ l_{n \Pi}\nabla^\mu \Pi
- l_{n \pi}\Delta^{\mu\nu} \partial_\gamma \pi^\gamma_\nu  \nonumber \\
&&\!\!\!\!
- \tau_{n \Pi} \Pi \dot u^\mu
- \tau_{n \pi}\pi^{\mu \nu} \dot u_\nu
+\lambda_{n\pi}n_\nu \pi^{\mu \nu}
+ \lambda_{n \Pi}\Pi n^\mu
+  \Lambda_{n \dot u} \omega^{\mu \nu} \dot u_\nu  \nonumber \\
&&\!\!\!\!
+ \Lambda_{n \omega} \Delta^\mu_\nu \partial_\gamma \omega^{\gamma \nu}
+ (9 \ {\rm terms}), \label{heat}\\
\pi^{\mu\nu} \!\!\!\!&=&\!\!\!\! \tilde a \pi_{\rm NS}^{\mu\nu} 
-\beta_{\dot \pi} \tau_\pi \dot \pi^{\langle \mu\nu\rangle}
+ \tau_{\pi n} n^{\langle\mu}\dot u^{\nu\rangle}
+ l_{\pi n} \nabla^{\langle \mu}n^{\nu\rangle}
+ \lambda_{\pi\pi} \pi_\rho^{\langle \mu} \omega ^{\nu\rangle \rho}  \nonumber\\
&&\!\!\!\!
- \lambda_{\pi n} n^{\langle\mu} \nabla^{\nu\rangle} \alpha
- \tau_{\pi\pi} \pi_\rho^{\langle\mu} \sigma^{\nu\rangle\rho}
- \delta_{\pi\pi} \pi^{\mu\nu}\theta
+ \Lambda_{\pi\dot u} \dot u^{\langle \mu} \dot u^{\nu\rangle}  \nonumber\\
&&\!\!\!\!
+ \Lambda_{\pi\omega} \omega_\rho^{\langle \mu} \omega^{\nu\rangle\rho}
+ \chi_1 \dot b_2 \pi^{\mu\nu}
+ \chi_2 \dot u^{\langle \mu} \nabla^{\nu\rangle} b_2
+ \chi_3 \nabla^{\langle \mu} \nabla^{\nu\rangle} b_2, \label{shear}
\end{eqnarray}
in the usual notation \cite{Jaiswal:2012qm}. The ``8 terms" (``9 
terms'') involve second-order, linear scalar (vector) combinations 
of derivatives of $b_1,b_2$. We observe that in the above equations, 
the nonlocal coefficients ($a,~b_1,~b_2$) modify the standard NS as 
well as the IS terms. This method is also able to generate all 
possible second-order terms allowed by symmetry.
\begin{figure}[t] 
\begin{center} 
\includegraphics[width=9.0cm, height=6.0cm]{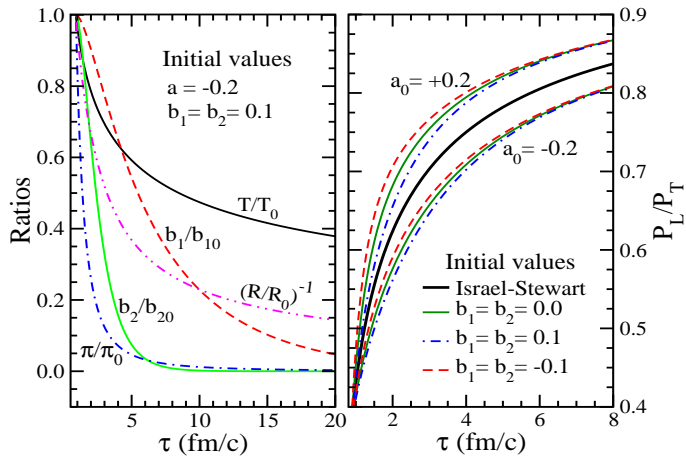}
\end{center} 
\vspace{-0.6cm}
\caption{Time evolution of temperature, shear pressure, inverse
  Reynolds number and parameters ($b_1, \ b_2$) normalized to their 
  initial values (left panel), and anisotropy parameter $P_L/P_T$ 
  (right panel). Initial values are $\tau_0= 0.9$ fm/c, $T_0=360$ 
  MeV, $\eta/s = 0.16$, $\pi_0 = 4\eta/(3\tau_0)$.}
\label{evolu}
\end{figure} 

\section{Results and discussion} 
To illustrate the numerical significance of the non-locality in the 
collision term through the new dissipative hydrodynamic equations 
derived here, we consider Bjorken evolution of a massless Boltzmann 
gas ($\epsilon=3P$) at vanishing net baryon number density 
\cite{Bjorken:1982qr}.

In terms of the coordinates ($\tau,x,y,\eta$) where $\tau =
\sqrt{t^2-z^2}$ and $\eta=\tanh^{-1}(z/t)$, the initial four-velocity
becomes $u^\mu=(1,0,0,0)$. In this scenario $\Pi=0=n^\mu$ and the
equation for $\pi \equiv -\tau^2 \pi^{\eta \eta}$ reduces to
\begin{eqnarray}\label{Bjpi}
\frac{\pi}{\tau_\pi} + \beta_{\dot\pi}\frac{d\pi}{d\tau} = 
\beta_\pi \frac{4}{3\tau} - \lambda \frac{\pi}{\tau} 
- \psi \pi \frac{db_2}{d\tau},
\end{eqnarray}
where the coefficients are
\begin{equation}\label{coeffu}
\beta_{\dot\pi} = \tilde a + \frac{4b_2\,P}{\tilde a \beta\eta},
~~ \beta_\pi = \frac{2}{3} \tilde a P ,
~~ \psi = \frac{22P}{3 \tilde a \beta \eta},
~~\lambda = 2\tilde a - 2\frac{b_1\beta^2 - 20 b_2}{3\tilde a \beta \eta} P.
\end{equation}
The coupled differential equations (\ref{evol}), (\ref{param}) and
(\ref{Bjpi}) are solved simultaneously for a variety of initial
conditions relevant for RHIC and LHC.

Figure \ref{evolu} (left panel), shows the evolution of several 
quantities for a particular choice of initial conditions. $T$ 
decreases monotonically to the crossover temperature $T_c\simeq 170$ 
MeV at time $\tau \simeq 10$ fm/c. Parameters $b_1$, $b_2$ vary 
smoothly to zero at large times indicating reduced but still 
significant presence of nonlocal effects at late times. This is also 
evident in Fig. \ref{evolu} (right panel), where the pressure 
anisotropy $P_L/P_T=(P-\pi)/(P+\pi/2)$ shows marked deviation from 
IS. Although the shear pressure $\pi$ vanishes rapidly indicating 
approach to ideal fluid dynamics, the $P_L/P_T$ is far from unity. 
Faster isotropization for initial $a>0$ may be attributed to a 
smaller effective shear viscosity in the modified NS equation.

\begin{figure}[t]\begin{center}
\includegraphics[width=7.5cm, height=5.8cm]{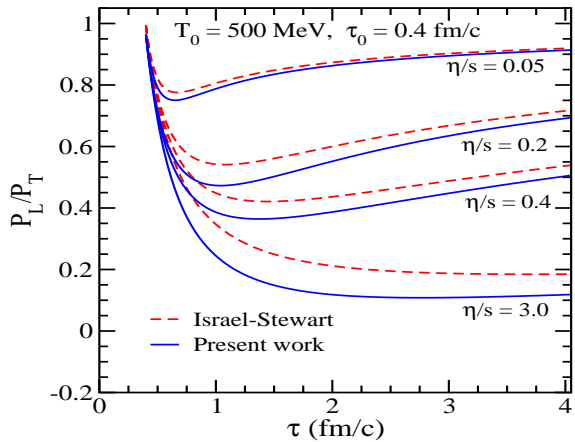}
\end{center}
\vspace{-0.6cm}
\caption{Time evolution of $P_L/P_T$ in IS (dashed lines) and the 
present work (solid lines) for isotropic initial pressure 
configuration.}
\label{evol2}\end{figure}

Figure \ref{evol2} shows the evolution of $P_L/P_T$ for isotropic 
initial pressure configuration, at various $\eta/s$ for the LHC 
energy regime. With small initial corrections ($\sim 10$\% to NS and 
$\sim 20$\% to the IS terms) due to $a,\ b_1,\ b_2$, nonlocal 
hydrodynamics (solid lines) exhibits appreciable deviation from the 
(local) IS theory (dashed lines). The above results clearly 
demonstrate the importance of the nonlocal effects, which should be 
incorporated in transport calculations as well.

\section{Summary}
To summarize, we have derived viscous hydrodynamic equations by 
introducing a nonlocal generalization of the collision term in the 
Boltzmann equation. The Navier-Stokes as well as Israel-Stewart 
equations are modified and new terms are obtained in the evolution 
equations of the dissipative quantities. The method presented is 
able to generate all possible terms that are allowed by symmetry. 
Within one-dimensional scaling expansion, we find that nonlocality of 
the collision term has a rather strong influence on the evolution of 
the viscous medium via hydrodynamic equations.

\end{document}